\documentclass[aps,twocolumn,showpacs,amsmath,amssymb,pre,superscriptaddress, floatfix]{revtex4-1}

\usepackage{graphicx}
\usepackage{dcolumn}
\usepackage{bm}
\usepackage{amssymb}
\usepackage{multirow}

\newcommand{\1}{\begin{equation}}
\newcommand{\2}{\end{equation}}
\newcommand{\ea}{\begin{eqnarray}} 
\newcommand{\ee}{\end{eqnarray}}

\newcommand{\Sum}[2]{{\sum\limits_{#1}^{#2}}}

\begin{document}
\title{Spatiotemporal Oscillation Patterns in the Collective Relaxation Dynamics of Interacting Particles in Periodic Potentials}

\date{\today}


\author{Benno Liebchen}
\email[]{Benno.Liebchen@physnet.uni-hamburg.de}
\affiliation{Zentrum f\"ur Optische Quantentechnologien, Universit\"at Hamburg, Luruper Chaussee 149, 22761 Hamburg, Germany}%
\author{Peter Schmelcher}
\email[]{Peter.Schmelcher@physnet.uni-hamburg.de}
\affiliation{Zentrum f\"ur Optische Quantentechnologien, Universit\"at Hamburg, Luruper Chaussee 149, 22761 Hamburg, Germany}%
\affiliation{The Hamburg Centre for Ultrafast Imaging, Universit\"at Hamburg, Luruper Chaussee 149, 22761 Hamburg, Germany}

\begin{abstract} 
We demonstrate the emergence of self-organized structures in the course of
the relaxation of an initially excited, dissipative and finite chain of interacting particles in a periodic potential
towards its many particle equilibrium configuration.
Specifically we observe a transition from an in phase correlated motion via phase randomized oscillations 
towards oscillations with a phase difference $\pi$ between adjacent particles thereby
yielding the growth of long time transient spatiotemporal oscillation patterns. 
Parameter modifications allow for designing these patterns, including steady states and even 
states that combine in phase and correlated out of phase oscillations along the chain. 
The complex relaxation dynamics is based on finite size effects together with an
evolution running from the nonlinear to the linear regime thereby
providing a highly unbalanced population of the center of mass and relative motion. 
\end{abstract}
\maketitle

\begin{figure}[htb]
\centering 
\includegraphics[width=0.48\textwidth]{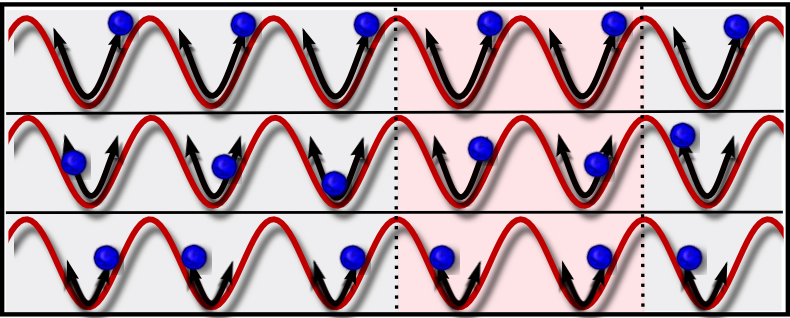}
\caption{\small (Color online) Schematic illustration of the setup, the initial state and its time evolution. On short timescales all particles oscillate in phase 
(upper panel), then
irregularly (middle panel) and on long timescales we have antiphase oscillations (lower panel). Dotted lines set the limits for a two particle system.}
\label{fig0}
\end{figure}

\begin{figure}[htb]
\centering 
\includegraphics[width=0.48\textwidth]{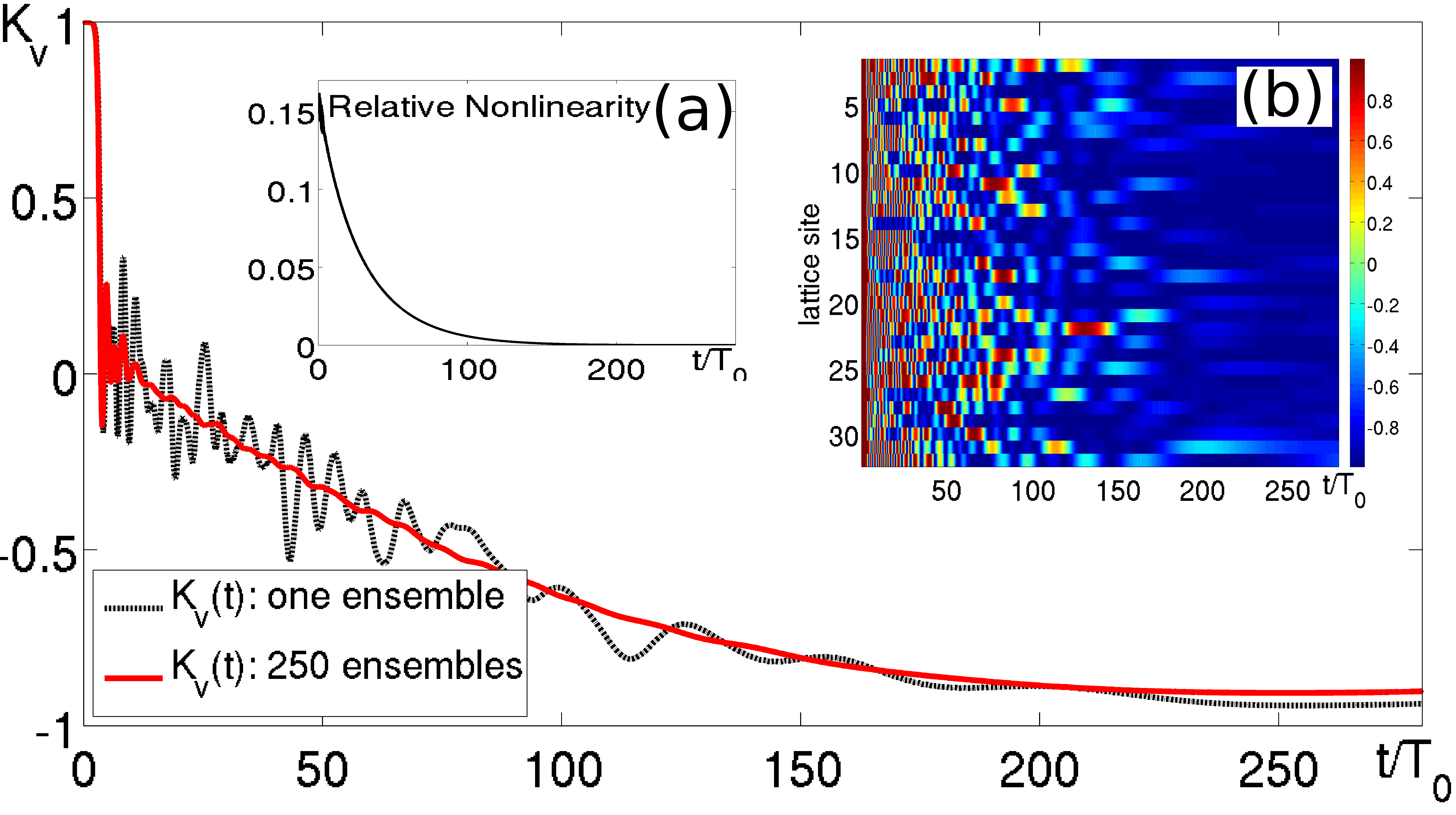}
\caption{\small (Color online) Time evolution of the velocity correlation function for a single ensemble (black, dotted) and averaged over
250 ensembles (red) with $x_0=1.0, v_0=0.1$. Inset (a): Time evolution of the relative difference 
of the lattice potential and its lowest order harmonic expansion $(1/N)\sum_j [\langle x_j^2 \rangle_{T_n} -\langle \sin(x_j)^2 \rangle_{T_n}]/\langle x_j^2 \rangle_{T_n}$. 
Inset (b): Time evolution of the site resolved velocity cross correlation function 
for a single ensemble (corresponding to the black, dotted curve) for parameters $\gamma=0.005,\alpha=r=1,N=32$.}
\label{fig1}
\end{figure}

\begin{figure}[htb]
\centering 
\includegraphics[width=0.48\textwidth]{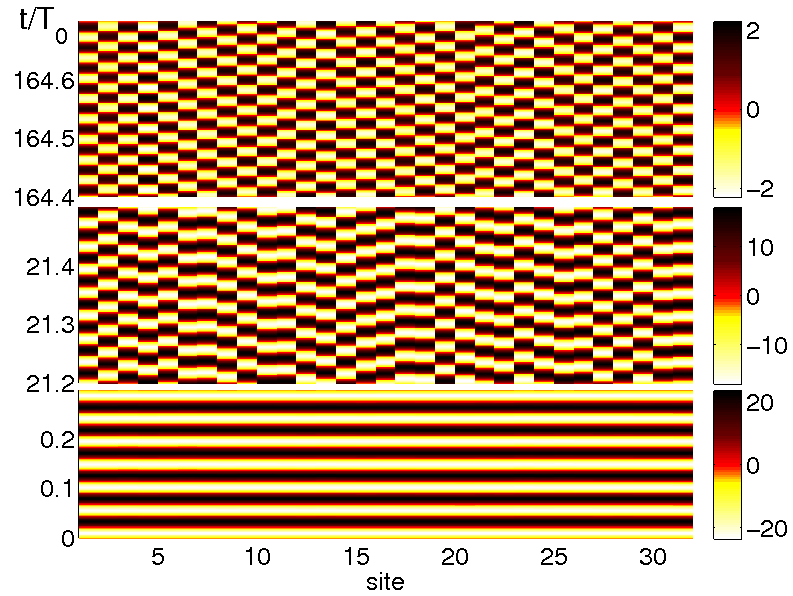}
\caption{\small (Color online) Extracts of the site resolved time evolution of velocity trajectories. Color/greyness provides the velocity values.
Parameters and initial conditions as in Fig.~\ref{fig1}.}
\label{fig2}
\end{figure}

\paragraph*{Introduction} 
Nonlinear dynamics is at the heart of the emergence of structure and complexity in nonequilibrium systems 
ranging from pattern formation in biological \cite{turing53,capasso13,cross93},
chemical \cite{turing53,maini97,antal99}
and physical systems \cite{cross93,cross09}
via the emergence of solitons, kinks and breathers in coupled nonlinear oscillators \cite{braun04} to the synchronization of self-sustained \cite{kuramoto84,pikovski01,acebron05} and chaotic oscillators \cite{pikovski96,pikovski01}. 
In view of the formidable progress achieved in recent years with respect to the cooling and trapping of particles \cite{pethick08,bloch08} and the control of their interactions \cite{chin10},
it is highly desirable to study the emergence of structure and complexity 
out of equilibrium with the extremely well controlled and prepared ensembles provided by cold atoms in optical lattices \cite{bloch08,lewenstein12} and ions in microtraps \cite{hughes11,wilpers12}.
Specifically for ions it is known that they possess already for their equilibrium a plethora of different configurations, such as zig-zag chains \cite{shimshoni11,mielenz13} and
ion crystals possessing concentric rings (2D), shells (3D) \cite{bonitz08} and 'string-of-disks' configurations \cite{kjaraard03}. Even 
two component Coulomb bicrystals exhibiting cylindrical structures coexisting with structures of spheroidal shape could be observed \cite{hornekar01}. 
Recent examples following the route of structure formation out of equilibrium 
include the pattern formation of trapped ions in an array of optical microtraps \cite{lee11}, the growth 
of density waves in driven superlattices \cite{petri11} and the emergence of dynamical current reversals associated with peaked velocity distributions for dilute long range interacting particles in driven lattices \cite{liebchen12}.
\\Here, we explore the complex pathway a highly excited, nonlinear chain of interacting particles takes in a dissipative periodic potential towards its asymptotic equilibrium configuration.  
We hereby demonstrate the emergence of a transition from initially in phase correlated motion via phase randomized oscillations towards oscillations with a phase difference $\pi$ between adjacent particles
('antiphase oscillations') yielding the growth of a long time transient spatiotemporal oscillation pattern. 
From the viewpoint of nonlinear dynamics this complex relaxation dynamics is demonstrated to be based on an interplay of finite size effects and a highly unbalanced population of the center of mass and relative motion
occurring in the course of the time evolution from the nonlinear to the linear regime.
Appropriate parameter modifications allow for steady state patterns and patterns of in phase and antiphase oscillations coexisting along the chain.
\paragraph*{Setup and Observables}
We consider a chain of $N$ point particles with coordinates $x_j$ and mass $m$ in a periodic lattice potential $V(x)=V_0\cos^2(k x)$ 
exposed to a frictional force $F_R=-\gamma \dot x$ and repulsive power law interactions.
The dynamics of this system is captured by the following $N$-dimensional system of coupled nonlinear Newtonian equations of motion
\1 m \ddot x_j + \gamma \dot x_j + V_0 k \sin(2 k x_j) - \Sum{k=1, k\neq j}{N} \frac{\alpha (x_j-x_k)}{|x_j-x_k|^{r+2}} = 0 \label{eqmo},\2
with $j=1,2..N$. We choose an initial ensemble with one particle per lattice site, a collective elongation $x_j(t=0)=(j+1/2)L + x_0$ with $L=\pi$ (compare Fig.~\ref{fig0}), 
and small but random initial velocities $\dot x_j(t=0)\in (-v_0,v_0)$ drawn from a uniform probability distribution.
Parameters are chosen according to the weakly damped ($\gamma T/m \ll 1$) and weakly interacting ($\alpha/(V_0 L^{r}) \ll 1$) regime throughout this article. Therein 
$T:=2\pi \sqrt{L^2 m/(2V_0)}$ is the typical timescale of one oscillation period in the lattice
and $V_0$ serves as the dominant energy scale such that each particle is captured on its initial lattice site in the course of the dynamics. 
Choosing natural space, time and mass units $(x_u,t_u,m_u)=(1/k,\gamma/(V_0 k^2),\gamma^2/V_0 k^2)$ the parameter space corresponding to Eq.~(\ref{eqmo}) reduces to two essential dimensions.
Thus, we use $m=k=1$ and $V_0=400$ without loss of generality and understand e.g. $\alpha,\gamma$ as effective parameters which can be adjusted via the other
parameters. 
\\In order to quantify the emergence of collective behavior and corresponding order in the lattice
in the course of the relaxation dynamics of the elongated particle chain,
we define the following velocity cross correlation function $K_v^{(i)}(t)$ providing a 
measure for the phase correlation of adjacent oscillating particles:
\1 K_v^{(i)}(t_n):=\frac{2\langle \dot x_i \cdot \dot x_{i+1}\rangle_{T_n}}{\langle \dot x_i^2\rangle_{T_n} + \langle \dot x^2_{i+1} \rangle_{T_n}} \label{velcor}\2
Therein $\langle.\rangle_{T_n}$ denotes the time average over the interval $[(n-1)T_0,n T_0)$ with length $T_0=2\pi \gg T$ and $t_n:=(n-1/2)T_0$. 
The corresponding ensemble average is denoted by $K_v(t_n)$. A system state yielding $K_v=1$ is referred to as phase correlated (PC) and a state yielding $K_v=-1$ is referred to as antiphase correlated (AC) in the following. 
Note that the corresponding position correlation function yields a similar behavior but includes contributions due to the inhomogeneity of the finite particle chain which renders it less appropriate for our present purpose.  
\paragraph*{Antiphase correlations and relaxation patterns}
We now explore the time evolution of the collectively elongated initial state by integrating Eqs.~(\ref{eqmo}) using the Dormand-Prince method. 
For very short times (see Fig.~\ref{fig1}) we observe $K_v \approx 1$ which corresponds to a PC
oscillation of the particle chain. Around $t/T_0 \sim 3$ the velocity correlation rapidly decays towards $K_v \sim 0$ and immediately afterwards starts to oscillate, even when considering the average of many ensembles (red curve in Fig.~\ref{fig1}). 
Consecutively, $K_v$ departs from a zero value and slowly decays towards $K_v \sim -1$.
This demonstrates the growth of antiphase correlations in the relaxation dynamics of the particle chain (compare also Fig.~\ref{fig1}b). 
It is illustrative to resolve this behavior of $K_v(t)$ also on the level of the corresponding velocity trajectories (Fig.~\ref{fig2}).
On short timescales (lower panel) the latter exhibit collective PC oscillations which then decay and yield a predominantly phase randomized intermediate dynamics (middle panel). 
For long times (upper panel) we have $K_v \sim -0.8$ which already corresponds to a well pronounced pattern of AC oscillations.
The remaining velocity amplitudes at this time are about 10 percent of the values reached within the first oscillation period after $t=0$. 
We note that the occurrence of the transition to AC oscillations occurs independently of the specific system size, for a wide range of parameters and is to some extend tunable via e.g. $V_0,\gamma$. 
\\The key question therefore is: Why does the considered particle chain not remain phase randomized but 
organizes itself into an AC oscillation pattern?
For the following reasons, a corresponding answer has to be developed from an essentially nonlinear many particle perspective which is also accounting for finite size effects of the system:
Even though the observed antiphase correlations seem to grow with decreasing nonlinear contributions to the lattice force 
(Fig.\ref{fig1}a), they would not occur in a corresponding system with linearized lattice force: 
In the latter case the motion of the center of mass (CM) $x:=(1/N)\sum x_i$ represents a normal mode that is decoupled from all other normal modes and obeys a damped single particle harmonic oscillator equation $\ddot x + \gamma \dot x + V x = 0$.
Since our collective initial state is constituted by the excitation of the CM mode only, its time evolution would exclusively lead to a damping of this mode.  
That means, the decay of the collective initial state and therefore also the emergence of antiphase correlations is crucially based on the nonlinearity of the lattice potential. 
Even more, for $v_0=0$ finite size effects are essential for the occurrence of the observed PC-AC-transition: in an infinite chain or a finite chain with periodic boundary conditions, 
translational invariance of the system and initial state yields an identical dynamics for all particles of the ensemble. 
\paragraph*{Analysis of the two particle chain}
In view of the complexity of the nonlinear power law interacting $N$-particle system we develop our physical understanding of the observed relaxation processes
starting from the underlying two particle system (compare Fig.~\ref{fig0}). 
We first demonstrate that in the respective Hamiltonian case ($\gamma = 0$) oscillations occur between pure center of mass motion (CMM) and pure relative motion (RM), that means between $K_v=1$ and $K_v=-1$ (compare Eq.~\ref{velcor}). 
Then, we deduce in the presence of dissipation, how these oscillations decay to result in $K_v \approx -1$, persistently.
\\Eq.~(\ref{eqmo}) reduces for $N=2$ ($m=k=r=1$) in elongation coordinates $x_j \mapsto x_j-(j+1/2)\pi$ to: 
\ea
\ddot x_1 + \gamma \dot x_1 + V_0 \sin(2x_1)+\alpha/(x_2-x_1+L)^{2} &=& 0  \nonumber \\
\ddot x_2 + \gamma \dot x_2 + V_0\sin(2x_2)-\alpha/(x_2-x_1+L)^{2} &=& 0 \label{2partsys}
\ee
\begin{figure}[htb]
\centering 
\includegraphics[width=0.48\textwidth]{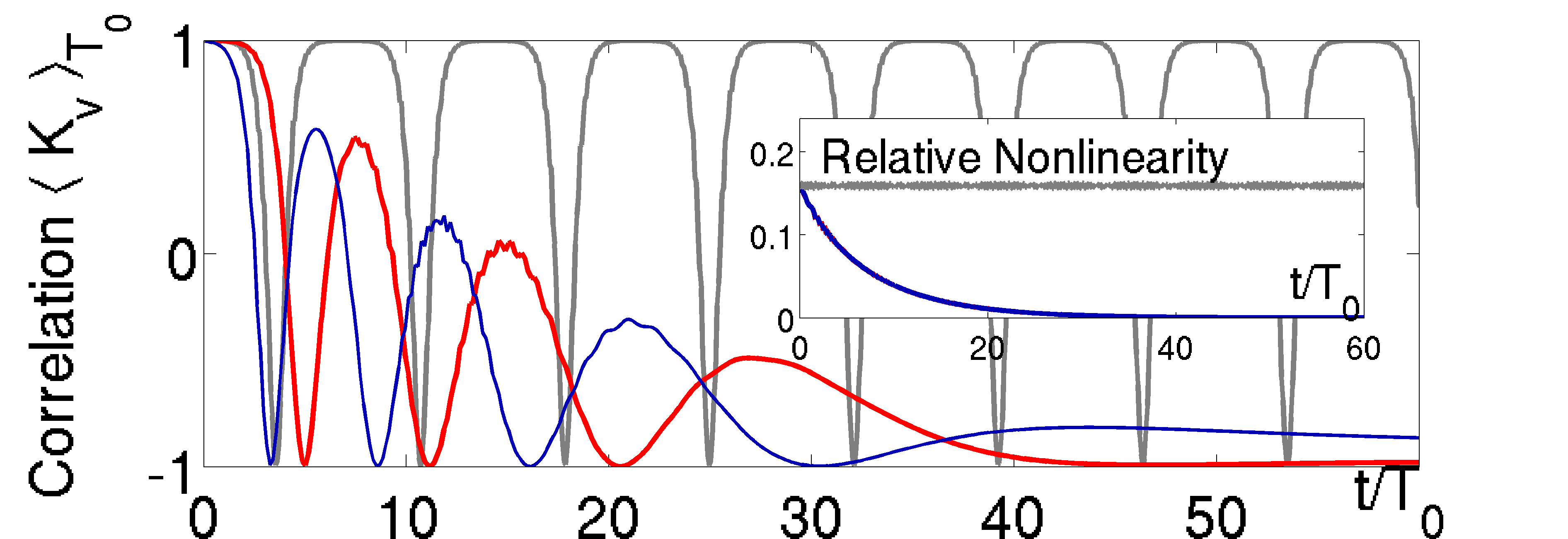}
\caption{\small (Color online) Time evolution of the velocity cross correlation function for the two particle system with $V_0=400$, $x_0=1$ and $\gamma=0; v_0=0$ (grey, low saturation), $\gamma=0.02; v_0=0$ (red) and
$\gamma=0.02; \dot x_1(t=0)=0.8; \dot x_2(t=0)=-0.54$ (blue). Inset:
Time evolution of the relative difference of the lattice potential and its lowest harmonic order expansion: $(1/N)\sum_j [\langle x_j^2 \rangle_{T_n} -\langle \sin(x_j)^2 \rangle_{T_n}]/\langle x_j^2 \rangle_{T_n}$ (the blue curve is on top of the red one).}
\label{fig3}
\end{figure}
Transforming Eqs.~(\ref{2partsys}) from elongation coordinates to the corresponding center-of-mass and relative coordinates $(x_1,x_2) \mapsto (X:=(x_1+x_2)/2,x:=(x_2-x_1))$ shows that the nonlinearity of the lattice
potential couples the CMM and the RM: 
\ea 
\ddot X + \gamma \dot X+ V_0 \sin(2X)\cos(x) &=& 0 \nonumber \\
\ddot x + \gamma \dot x + 2 V_0 \cos(2X) \sin(x) - 2\alpha/(x + L)^2 &=& 0. \label{2partcofrel}
\ee  
The initial oscillations of the chain yield $K_v=1$, i.e. the total kinetic energy is in the CMM which corresponds to a phase correlated oscillation of both particles. 
The coupling between the CMM and the RM then causes a flow of the kinetic energy from the CMM to the RM and we observe a 
corresponding time evolution of $K_v$ from $1$ to $-1$ in Fig.~\ref{fig3}. 
Then, the kinetic energy flows back to the CMM, $K_v(t)$ increases from $-1$ to $1$ and subsequently
possesses approximately periodic oscillations, which is expected by the KAM-theorem since  
Eqs.~(\ref{2partsys}) reduce for $\alpha=\gamma=0$ to two conservative and independent single degree of freedom systems (physical pendula) 
which are integrable, and the weak coupling $\alpha/L^r \ll V_0$ yields only a weak perturbation of this integrable case. 
In the presence of dissipation ($\gamma>0$), with decreasing energy and elongations $x_1,x_2$, we obtain a transition from coupled nonlinear oscillators to coupled linear oscillators. 
While in the nonlinear regime there occur periodically transitions between pure CMM and pure RM (Fig.~\ref{fig3}), 
the latter decouple in the linear regime (for long times) with occupations inherited from the nonlinear history. 
\\Let us now address the question why the dynamics of the population of the CMM and the RM converges for long times, independently of the specific initial 
velocities, to a pure occupation of the RM corresponding to $K_v \sim -1$ for $t/T_0 \gg 1$ (see red and blue curve in Fig.~\ref{fig3}). 
Using the Rosenberg-scheme \cite{rosenberg62} a nonlinear normal mode (NNM) can be detected to exist for the full two particle system Eq.~(\ref{2partsys}) which reads $x_1(t)=-x_2(t)$ and expresses the mirror symmetry of the two particle system.
Accordingly, for initial conditions $x_1(t=0)=-x_2(t=0), \dot x_1(t=0)=-\dot x_2(t=0)$, Eqs.~(\ref{2partsys}) reduce to the effectively one degree of freedom problem 
$\ddot x_1 + \gamma \dot x_1 + V_0\sin(2x_1)+ \alpha/(2x_1-\pi)^2, x_2(t)=-x_1(t)$.  
It can be numerically shown, that the NNM is stable with respect to perturbations of the velocities (if and only if $\alpha>0$),
i.e. a configuration $x_1(t_0)=-x_2(t_0)$ yields trajectories $x_1(t) \approx -x_2(t)$ (RM)
if $\dot x_1(t_0) \approx -\dot x_2(t_0)$ holds.
In contrast, a configuration $x_1(t_0)=x_2(t_0)$ always leads to a combination of CMM and RM, even for $\dot x_1(t_0)=\dot x_2(t_0)$.
Since in the course of the dynamics of the dissipative two particle chain, consecutive pure populations of the RM 
(consecutive minima in Fig.~\ref{fig3}) yield lower and lower particle velocities
the flow of the population from the RM to the CMM is increasingly suppressed by the NNM, while the opposite flow is unhindered. 
Thus, we observe that $K_v(t)$ overall approaches the value $-1$ for long times in Fig.~\ref{fig3}. 
\paragraph*{$N$-particle chain} 
Let us now transfer our understanding of the crossover from $K_v(t)\sim 1$ to $K_v(t)\sim -1$ for the two particle chain to  
the relaxation dynamics of the complex $N$-particle chain.  
\begin{figure}[htb]
\centering 
\includegraphics[width=0.48\textwidth]{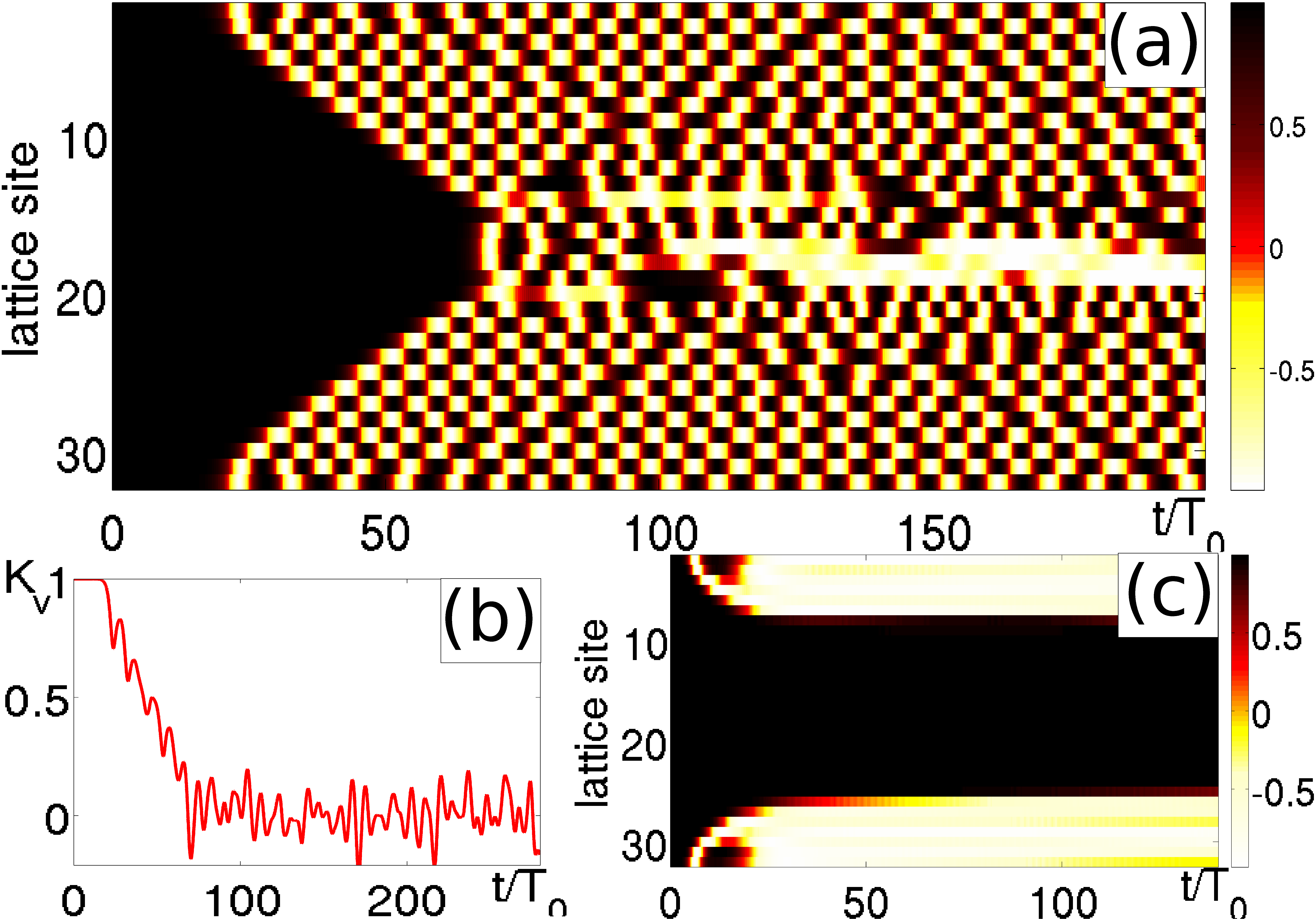}
\caption{\small (Color online) Time evolution of the site resolved (a,c) and the ensemble averaged (b) velocity cross correlation function. Parameters: $\gamma=0$, $\alpha=0.04, N=32,r=1$ (a,b) and
$\alpha=r=1,N=32,\gamma=0.029$ (c).}
\label{fig4}
\end{figure}
We first discuss the asymptotically emerging anticorrelations ($K_v \sim -1$) as observed in Figs.~\ref{fig1},\ref{fig2} 
and then analyze the decay of the initial CMM of the chain (see Fig.~\ref{fig1}).
Like in the two particle case, the RM $x_j(t)=-x_{j+1}(t);\; j=1,2..N-1$ is a NNM of the $N$-particle chain (compare Eqs.~\ref{eqmo}).
Thus, whenever the $N$-particle chain reaches a state $x_j \approx -x_{j+1}; \; \dot x_j \approx - \dot x_{j+1} \; \forall j$ at some point in time, 
the population of the RM is approximately preserved by the stable NNM. 
That means, like in the two particle case, the coupling between the CMM and the RM by the nonlinear contributions of the lattice potential 
allows for a population of the RM in the nonlinear regime, but its 
depopulation is increasingly prohibited by the NNM the closer the dissipative chain approaches the linear regime.    
\\Let us now discuss the finite size induced decay of the initial CM oscillation of the chain, which is best resolved in the 
case of vanishing initial velocities $\dot x_j(t=0)=0$
and in the absence of a decay of nonlinear terms ($\gamma=0$): 
In Fig.~\ref{fig4}a we observe a checkerboard-like pattern of spatiotemporally alternating phases of 
correlated and anticorrelated oscillations emerging from the initial CMM of the particle chain.   
The growth of this pattern from the edges of the chain towards its center can be understood due to finite size effects: 
Initially, when the chain exhibits a pure CM motion, the total interaction forces acting on particles in the center of the chain are 
similar to the forces acting in a corresponding translationally invariant infinite chain which preserves for $v_0=0$ for all times
the initial CMM of the chain. In contrast, the total interaction forces acting on particles at the two edges of the finite chain are comparatively large
and yield, as in the two particle case, a RM between the outermost particles and the remaining chain which performs its initial CMM.
Once, the outermost particles are out of phase with respect to the rest of the chain, the effective interaction forces acting on the second outermost particles enhance
and evoke a corresponding relative motion between these particles and the remaining part of the chain performing CM oscillations. 
This scenario repeats until the CM oscillations of the whole chain are dissolved. 
Obviously the checkerboard pattern is at no time perfect (Fig.~\ref{fig4}a). Correspondingly, the $N$-particle chain is subject to irregularly fluctuating 
interaction forces, which do not decay for $\gamma=0$ and
yield a slowly emerging phase randomization (Figs.~\ref{fig4}a,b).
\\A similar checkerboard-like pattern as in Fig.~\ref{fig4}a emerges also for $\gamma>0$ on short timescales (not shown) and 
even if we additionally have small initial velocities ($v_0>0$) (see Fig.~\ref{fig2}b) some signatures of this pattern remain.  
We note that the occurrence of antiphase correlations (Fig.~\ref{fig1}) emerges together with a shrinking of the
$2N$-dimensional phase space, first towards the $N$-dimensional subspace of pure RM ($x_i(t)=-x_{i+1}(t)$) and 
asymptotically onto the one dimensional fixed point attractor corresponding to the $N$-particle equilibrium configuration.
\paragraph*{Steady state and coexisting dual patterns}
Let us now address the issue of tunability and parameter dependencies.
Frequency and lifetime of the PC-AC-oscillations as observed in Fig.~\ref{fig4}a can be tuned with both, the interaction 
strength which determines the asymmetry of the interactions in the finite chain, and 
the initial coupling between CMM and RM, i.e. via the initial elongation $x_0$ of the particle chain.
We note that the observed oscillation patterns of the many particle chain persist for other power law interactions ($r \neq 1$). 
\\Modulations of $\gamma$ have a particularly interesting impact on the observed oscillation patterns. 
Firstly, once the AC pattern (Fig.~\ref{fig2}) is established for $\gamma=0.005$, a $\gamma \rightarrow 0$-quench allows to convert it
into a steady state pattern.
Secondly for specific values of the time independent dissipation coefficient a chimera-like dual pattern emerges that consists 
of a persistently phase correlated core of the chain coexisting with uncorrelated outer parts aspiring anticorrelations (Fig.~\ref{fig4}c).
In order to create such a dual pattern, we exploit the fact that for $v_0=0$ the initial CMM of the chain dissolves consecutively from the 
edges of the chain towards its center
(Fig.~\ref{fig4}a). While a sufficiently strong dissipation leads to a rapid decay of the impact of nonlinear terms 
of the lattice potential on the dynamics of the chain, an appropriate intermediate dissipation 
allows for the emergence of the checkerboard-like pattern at the edges of the chain (Fig.~\ref{fig4}a) 
but leads to a decay of the impact of nonlinear terms before the PC oscillations of the inner part of the chain are dissolved. 
Then, the inner part of the chain approaches the many particle steady state via its initial PC oscillation, while the outer regions of the chain aspire ACs.
\paragraph*{Conclusions}
The transition from the PC initial state via the phase randomized intermediate oscillations to the AC oscillations,
the checkerboard pattern occurring in the site resolved velocity correlation function 
and also the coexistence of PC and AC oscillations along the particle chain should be directly accessible 
in state of the art experiments with e.g. interacting colloidal particles in optical lattices \cite{bohlein12a,bohlein12} or 
with cold ions either in an array of microtraps \cite{hughes11,wilpers12,enderlein12} or in a segmented rf-trap with many dc-electrodes \cite{schulz08}. 
Using ${}^{40}$Ca${}^+$ ions and employing a distance of $L \sim 100\mu m$ between individual trapping regions as well as an oscillation frequency of $1/T \sim 10^{5}/s$ (i.e. $\alpha/(V_0 L) \sim 10^{-3}$), 
the latter allows for observing the emergence of anticorrelations on the timescale of $10^4$ oscillations of the particles in the lattice, assumed that the 
dissipation as provided by e.g. Doppler cooling or resistive cooling \cite{schulz08,itano95} 
is tuned to $\gamma \sim 10^{-23}kg/s$, i.e. $\gamma T/m \sim 10^{-3}$. 
Note that there is a formal analogy of our setup and an array of macroscopic classical pendula with ionic, macroscopic particle clusters attached, which
could allow even for a mechanical implementation. 
Our work might be useful in order to load lattices with various oscillation patterns and represents a first step 
in order to design and control the collective dynamics of excitations in complex systems on periodic substrates. 
The possibility to employ different masses, charges and initial states for our setup and to extend it to higher dimensions should provide a rich perspective in order to design oscillation patterns on the level of different observables including correlation functions. 

\end{document}